# Iterative Retina for high track multiplicity in a barrel-shape tracker and high magnetic field

W. Deng, Z. Song, G. Huang, G. De Lentdecker, F. Robert, Y. Yang

*Abstract*— Real-time track tracking in high energy physics experiments at colliders running at high luminosity is very challenging for trigger systems. To perform pattern-recognition and track fitting in online trigger system, the artificial Retina algorithm has been introduced in the field. Retina can be implemented in the state of the art FPGA devices. Our developments use Retina in an iterative way to identify track for barrel-shape tracker embedded in a high magnetic field and with high track multiplicity. As a benchmark we simulate LHC t-tbar events, with a pile-up of 200 and a GEANT-4 based simulation of a 6-layers barrel tracker detector made of silicon modules. With this sample the performance of the hardware design (resource usage, latency) is evaluated. Both efficiency and purity of the Retina fitting are over 90%. Moreover we have also added a Kalman filter after the Retina fit to improve the resolution on the track parameters. Our simulation results show that the Kalman filter can work well together with the Retina algorithm to find track through t-tbar event and provides high resolutions of the reconstructed parameters.

## I. Introduction

The research of fast tracking in online trigger system for the high energy physics experiments in collider running at high luminosity is very important. The purpose of fast online tracking is to identify the tracks and reconstruct the parameters of each track we are interested amount the collision event. It can not only save the bandwidth of data transmission between front-end readout system and offline computing center, but also provide preprocessing support for offline data analysis system to reduce the calculation steps of offline data processing.

Our research work is mainly aimed at fitting and reconstructing curved tracks from a barrel-shape tracker detector surrounded by a high magnetic field (typically 4 T). We introduced the retina algorithm as the main trajectory matching algorithm for fast tracking and used generated LHC t-tbar data samples and simulated with GEANT4 [1] the response of a 6-layer silicon barrel tracker. After getting promising results from simulation, we implemented it into FPGA devices to test its online performance (such as the resource usage and latency of the Retina algorithm).

In previous studies, we have designed a general Retina algorithm to reconstruct single muon tracks in a similar tracker geometry [2]. The strategy is first using the weight calculation formula of Retina algorithm to scan the parameter space and then find the most possible cells with maximum weight value as the result output of the algorithm. In that case, only one track can be identified after each operation of Retina algorithm.

In the present paper, we mainly introduce our research work about reconstructing the charged particle's trajectory in the strong magnetic field and high track multiplicity environment by using iteratively the Retina algorithm under the structure of barrel-like shape detector. Compared with previous research work, we have upgraded our algorithm to cope with more complex physical events, namely LHC t-tbar events with a pile-up of 200 collisions. Firstly, we have changed our strategy and used the iterative retina algorithm to scan the parameter space several times, so that the retinal algorithm can reconstruct multiple tracks at the same time. Secondly, we have added the Kalman filter after the Iterative Retina to improve the resolution of parameters reconstructed for each track found by the Iterative Retina.

We will introduce these features in four parts in detail:
 -The description of Retina and Iterative Retina algorithms
 -The introduction of the tracker environment and the configuration of Iterative Retina in our research
  -The firmware design of the Iterative Retina
  -The Kalman filter for track reconstruction

Besides, we have added an analysis part and discuss the result of our design for fast tracking (both the results come from the Iterative Retina and the Kalman filter) in the end.

## II. Retina and Iterative Retina for track fitting

The Retina algorithm is inspired from the processing of visual images by the brain where each neuron is sensitive to a small region of the retina. The strength of each neuron is proportional to how close the actual image projected on the retina region is to the particular shape that particular neuron is tuned to [3,4].

Manuscript submitted Oct 31, 2020. This work was supported in part by Inter-University Institute For High Energies, Université libre de Bruxelles (ULB), and the National Key Research and Development Program of China under Grant No. 2016YFE0100900

W. Deng is with both Center China Normal University (CCNU) and Université libre de Bruxelles (ULB) (E-mail: dengdandan@mails.ccnu.edu.cn)
Z. Song is with Naval University of Engineering (E-mail: Zixuan.Song@ulb.ac.be)
G. Huang is with both Center China Normal University (CCNU)

G. De Lentdecker, F. Robert and Y. Yang are with Université libre de Bruxelles (ULB)

When it used as the pattern recognition algorithm for track fitting, Retina algorithm firstly to defines a parameter space with several parameters need to be constructed, each parameter represents an attribute of the pattern to be recognized. Secondly to divide the parameter space into a group of cells with appropriate granularity, each cell represents a specific pattern. Then, each input data set is used to scan all cells in parameter space, and the cells whose output weight exceeds the threshold value are regarded as the corresponding patterns found by the Retina algorithm.

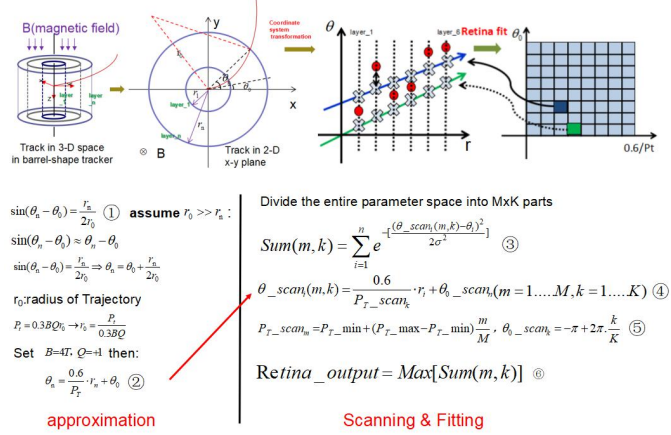

Fig. 1. The principle of using the Retina algorithm to fit and reconstruct 2-D parameters for curved tracks in a barrel-shape tracker detector.

The Figure 1 shows the principle of the Retina algorithm to fit and reconstruct curved tracks in a barrel-shape tracker with high magnetic field around. The cartoon at the top left corner of Figure 1 shows the ideal geometry of tracker detector; the full tracker is consisting of n concentric barrel-shape sensitive layers (named layer_1 to layer_n) and the direction of the magnetic field B is perpendicular to the x-y plane. All the simulated particles start from the center of the circles, with a given initial angle $\theta_0$, and cross the n layers from the center outwards. In that case, the Retina algorithm focus on the x-y plane to identify the 2-D pattern of curved tracks. So the 2-D parameter space of Retina algorithm is defined by two parameters, the initial angle of each tracks, $\theta_0$, and the transverse momentum of each particle, Pt. The Retina inputs are the hit coordinates $(r_n,\theta_n)$ on $n^{th}$ layer in the transverse x-y plane. There are two steps to complete the computation of Retina algorithm (shown in Figure. 1), first we do an approximation by linearizing the particle trajectory equation in the transverse plane: Eq. 1 -> Eq. 2. Secondly we divide the parameter space (0.6/Pt, $\theta_0$) into M*K cells (bins), for example one of those cells is (m,k), and scan the full space to find the possible track cell location who's output value is over the threshold we set according to formula ⑥. Where the Eq.2 in Figure. 1 is the final formula to present the relationship between the coordinate location $(r_n,\theta_n)$ in $n^{th}$ layer with the $\theta_0$ and Pt of the track pattern. The formula ③④ and ⑤ are the intermediate calculation for scanning, where the Sum(m,k) is the Retina weight output of parameter cell (m,k), the $(r_i,\theta_i)$ is the input hit location of the track in $i^{th}$ layer (i from 1 to n layer), the $\theta\_scan_i(m,k)$ is the angle location output by cell (m,k) in $i^{th}$ layer, the Pt_scan$_m$ and the $\theta_0\_scan_k$ are representing the parameters of the Pt and the initial angle for cell (m,k) respectively. Besides, the B stands for the strength of the magnetic field (the unit is Tesla(T)), the Q stands for the charge of the particles (the unit is the charge per unit electron) and the $r_0$ stands for the radius of the particle's trajectory.

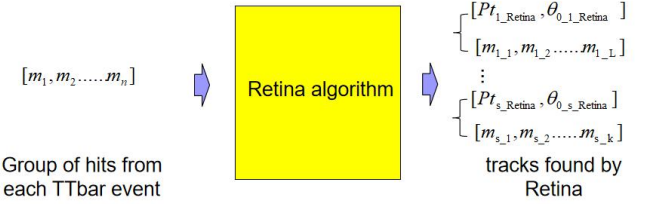

Fig. 2. The diagram to show the input and output of Retina algorithm for track fitting application in tracker detector.

Figure 2 summarizes the data flow through the Retina algorithm. At each event, a group of hits, $m_1$-$m_n$, are loaded into the Retina algorithm which will distinguish the various trajectories as explained in Figure 1, group the hits from each trajectory together (as show in Figure.2 $m_{1\_1}$ to $m_{1\_L}$ for $1^{th}$ track and $m_{s\_1}$ to $m_{s\_k}$ for $s^{th}$ track), and give the corresponding initial angle $\theta_0$ and transverse momentum Pt ($[Pt_{1\_Retina},\theta_{0\_1\_Retina}]$ for the $1^{th}$ track and $[Pt_{s\_Retina},\theta_{0\_s\_Retina}]$ for the $s^{th}$ track) in the end.

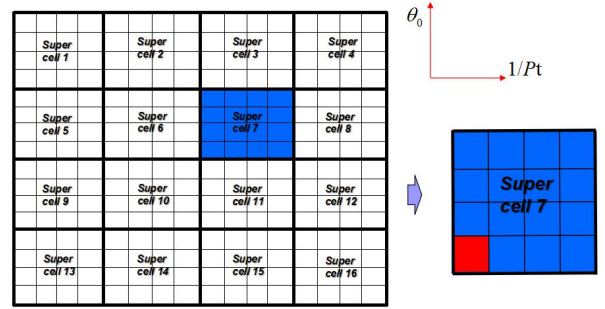

Fig. 3. Example of two iterations of 4bins($\theta_0$)*4bins(1/Pt) Retina algorithm with different granularity to cover 16bins($\theta_0$)*16bins(1/Pt) parameter space cells with highest granularity.

According to the principle of the Retina algorithm, if we need to get a relatively high resolution for the reconstructed parameters $\theta_0$ and Pt, we need to scan the parameter space with higher granularity. Obviously the higher granularity, the larger number of cells have to be scanned with Retina for each event. When we use the Retina algorithm to identify multiple trajectories in the case of the complex t-tbar event, it will cost huge time and resource to implement it into software or firmware. To avoid to scan all the parameter space cells with the highest granularity, we iteratively run Retina increasing the parameter space granularity at each step but scanning a limited number of parameter cells around. We called it Iterative Retina.

Figure.3 shows an example where we run 2 times Retina on 4*4 cells to reach the equivalent precision of a 16*16 granularity in the same parameter space. The gain in processing is the following: for a granularity of the parameter space of n x n with traditional Retina, one would have to scan n^2 cells while with the iterative way one has to scan, with i iterations, i * m^2 cells, where n = m^i. This means decreasing the number of cells to scan from n^2=m^(2i) to i*m^2. At each step the region of interest is defined as the cell with the Retina output value which is over the threshold.

So by using Iterative Retina, the computation and resource consumption of the algorithms can be reduced a lot when implementing both in software and firmware. Therefore, we decided to use Iterative Retina as the main fitting and reconstructing algorithm for the case of high track multiplicity.

## III. TRACKER SEGMENTATION AND ITERATIVE RETINA CONFIGURATION FOR IMPLEMENTATION

Our R&D work focuses on the development of fast tracking algorithms to be run on FPGA and we use the CMS Phase 2 outer tracker detector and fully simulated t-tbar events within the LHC CMS experiment as a benchmark for our developments. At this stage our results do not represent the most up-to-date CMS track trigger performance [5, 6].

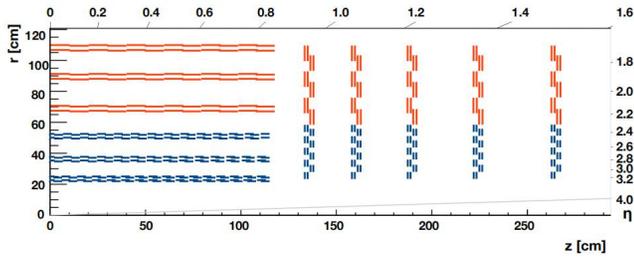

Fig. 4. The flat barrel design of one quadrant in the r − z plane of a proposed CMS Phase 2 Outer Tracker layout.showing the 2S (red) and PS(navy) module placement.

Figure 4 shows the one quadrant of the sensitive silicon detector layer of CMS Phase 2 outer tracker in r-z plane, we aimed at the η region from -0.6 to 0.6 where the tracker consists of 6 concentric barrel-shape sensitive layers. The radius of those 6 layers are from $r_1$=20cm (innermost) to $r_6$=115cm (outermost). The inner three layers are consist of PS modules, and the outer three layers are consist of 2S modules [7]. The direction of magnetic field is along the z axis and the strength is 3.8 Tesla.

There are four parameters to describe a track in the tracker region we are interested: the vertex position along z axis $z_0$; the initial angle of track in x-y plane $\theta_0$; the parameter η stands for the pseudo-rapidity; the Pt of the charged particles (the momentum in x-y plane). We plan to use Iterative Retina to reconstruct the first two parameters Pt and $\theta_0$, then apply the kalman filter to reconstruct all four parameters with higher resolution. In calculation, the unit of $\theta_0$ is centi radian (crad), the unit of Pt is GeV/c and the unit of $z_0$ is centimeter (cm).

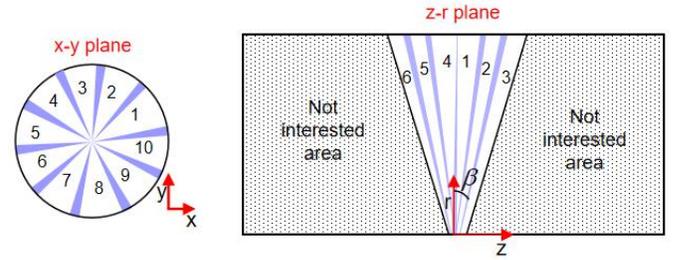

Fig. 5. The segmentation of the tracker for Iterative Retina implementation.

In order to implement the Iterative Retina with a resource limited FPGA device, we segment the tracker region we are interested into 60 sectors, 10 parts along direction in x-y plane and 6 parts along η direction in r-z plane. Figure 5 shows the segmentation of the tracker. The numbered areas in white represent the regions that are associated with only one sector, whereas the coloured areas represent the overlap region between neighbouring sectors where stubs may need to be assigned to both sectors. For each sector we use one processor to implement the Iterative Retina, where the processor is the MCU in simulation or the FPGA devices in firmware. Therefore, in one processor of Iterative Retina the scan range of parameter space for $\theta_0$ is one-tenth of full parameter space (-π,π) and the scan range of parameter space for Pt is >= 2Gev/c.

So in the end the scale of Iterative retina is adjusted to handle one sector. In that case, we present 3 configurations of Iterative Retina with different scan granularity. The first configuration consists in two iterations of Retina algorithm with the same scan granularity of 6 bins*8 bins, with 6 bins for the $\theta_0$ dimension and 8 bins for the Pt dimension, to reach the scan granularity of 36 bins*64 bins. In the second configuration we increase the scan granularity in both iterations from 6 bins to 8 bins for the $\theta_0$ and 8 bins to 10 bins for the Pt to reach the scan granularity of 64 bins*100 bins. In the third configuration, we use the highest scan granularity in both iterations with 10 bins for $\theta_0$ and 20 bins for Pt to reach the scan granularity of 100 bins*400 bins. We name those 3 configurations as Loop-48 Loop-80 and Loop-200 respectively.

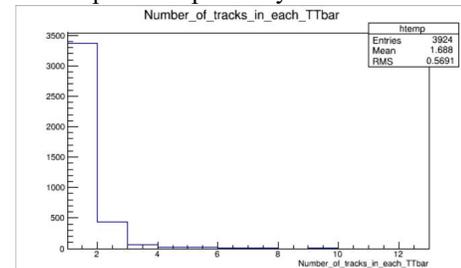

Fig. 6. The distribution of number of tracks in t-tbar event at one sector.

Besides，we use the fully simulated t-tbar event samples with a pile-up of 200 collisions. Before we input the data into our Iterative Retina processor, we filter the data and select only the hits which belong to one of the sector described above. In that case, we count the number of trajectories to be reconstructed in one sector for t-tbar event samples. Figure 6 shows the distribution of number of tracks reconstructed per the t-tbar event within one tracker sector . From the Figure 6, we can see almost 99% of t-tbar event that there are less than 3 tracks need to be identified. So in firmware design we fixed our scale of Iterative algorithm to deal with maximum 3 tracks for one process. And in simulation software there are no that limitations.

## IV. THE FIRMWARE DESIGN OF THE ITERATIVE RETINA

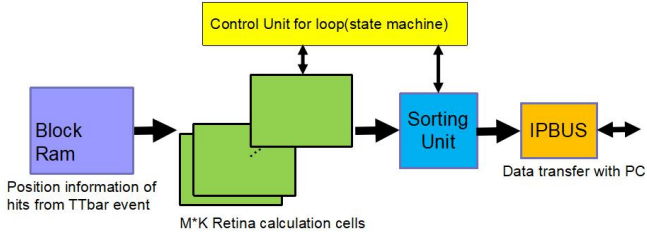

Fig. 7. The top FW design block diagram of Iterative Retina

After getting promising results from simulation，we then implement the Iterative Retina algorithm into FPGA devices to test its online performance. Figure 7 shows the top-level firmware design of Iterative Retina algorithm, it include 5 units：

The first unit is a block ram unit to store and refresh the position information of hits from t-tbar event as the input data for Iterative Retina.

The second unit is the array of Retina calculation cells that is used to scan the parameter space with the configurable granularity (M*K, M bins for $\theta_0$ and K bins for Pt) in a parallel way and output the weight value Sum(m,k) (m from 1 to M ,k from 1 to K) for all cells at the same time. According to the selected loop-configuration, there are 48 or 80 or 200 calculation cells in this unit respectively.

The third unit is the Sorting unit, it handles two functions: the first function is to pick up the super cells whom output value Sum(m.k) is over the threshold in the first iteration of the Retina algorithm; the second function is to select the cell with maximum output value Max(Sum(m.k)) in the second iteration of the Retina algorithm.

The fourth unit is a control unit consisting of a set of state machine. The mainly task of the control unit is to handle the work flow of the Retina iterations. At the beginning, the control unit will send the fixed configuration information to all Retina calculation cells in the Retina calculation array unit and start the first iteration of Retina algorithm. The configuration information includes the scan step length and scan location coordinate for each cells. After the first iteration of Retina algorithm is finished, the control unit receives the results output from the Sorting unit; the results are the location information for those super cells picked by the first iteration of Retina, and resend the refreshed configuration and process the second iteration of Retina algorithm for N times. The number N is equal to the number of super cells found by the first iteration of Retina algorithm. When both iterations of retina are completed, the control unit will input the data of the next t-tbar event and restart the program.

The last unit is the IPBUS [8] communication unit. It sends the results data from Retina block to PC for analysis. The bandwidth of IPBUS is up to 1Gbit/s, it is suitable for current firmware scale. Once we have lager firmware scale we can change from IPBUS to PCIE.

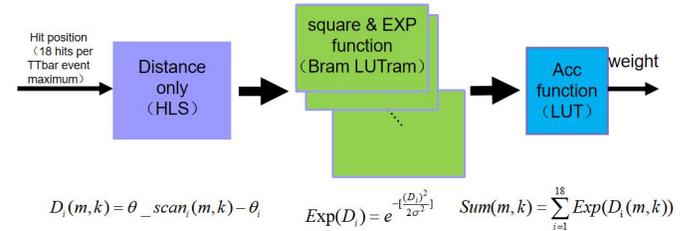

$$D_i(m,k) = \theta\_scan_i(m,k) - \theta_i \quad Exp(D_i) = e^{-\frac{(D_i)^2}{2\sigma^2}} \quad Sum(m,k) = \sum_{i=1}^{18} Exp(D_i(m,k))$$

Fig. 8. Block diagram of one Retina calculation cell

Inside one calculation cell, for example cell (m,k), we separate the Retina computation into three steps (see in Figure 8): First to calculate the distance $D_i$ between the input hit position $\theta_i$ and the angle location of cell (m,k) $\theta\_scan_i(m,k)$, this function is designed by the high level synthesis (HLS) tools of the Xilinx vivado toolkit. Then we make the square and exponential operation (green blocks in Figure 7 named EXP below) using the distance $D_i$. In those steps there are only one input and one output for the calculation, so we use look up table (LUT) resources in FPGA to implement this function and both block ram and distribute ram are suitable for it. Besides, all the hits in one event are processed in parallel in these steps. Therefore there are 18 LUT tables in each Retina calculation cell in firmware design. At last we have an accumulator to complete the Sum function, add all 18 results output by the EXP units as the weight output of one calculation cells.

TABLE I
THE RESOURCE USAGE AND LATENCY OF ITERATIVE RETINA IMPLEMENT IN FPGA DEVICE.

| Clock (Hz) | Firmware Design | DSP | LUT | BRAM | FF | Latency (Cycles)/μs |
|---|---|---|---|---|---|---|
| -------- | KC705 | 840 | 203800 | 445 | 407600 | -------- |
| 200M | One calculation cell | 1 | 441 | 0 | 517 | -------- |
| 200M | LOOP_200 | 200 (23.81%) | 112326 (55.2%) | 0.5 (0.11%) | 58289 (14.3%) | 312/1.56 |

The firmware design is already test in KC705 evaluation board [7], where an Kintex7 series FPGA chip are used. The latency and resource usage of the Iterative Retina are shown in Table 1. To process one event, it takes 312 clock cycles at the clock frequency of 200 MHz and it costs nearly half of the KC705 FPGA resources to compute 200 cells with Retina. And for each cell it costs 1 DSP and 441 LUT unit in FPGA device. So from the Table 1, we can see the current online performance, where we have the latency of 1.56 μs, is good enough to meet most of the LHC experiments hardware trigger requirement.

## V. KALMAN FILTER FOR TRACK PARAMETER RECONSTRUCTION

Considering of the reconstruction resolution requirement for LHC tracker trigger systems, typically a few % on the Pt, it is hard to meet this requirement with only the Iterative Retina although it has a good performance in efficiency and purity. Because in Iterative Retina we scan the 1/Pt instead of the Pt parameter, in case of high Pt (>10GeV/c) the resolution of Pt will decrease a lot. This is caused by the principle of pattern recognition, when the Pt of the particle is high enough the r0 of the track infinity approaches positive infinity and the shape of the trajectory is infinitely close to a straight line. So for pattern recognition algorithm (indeed Retina is belong to recognition algorithm) it is hardly to distinguish the curve shape of infinitely close to straight line from one to another.

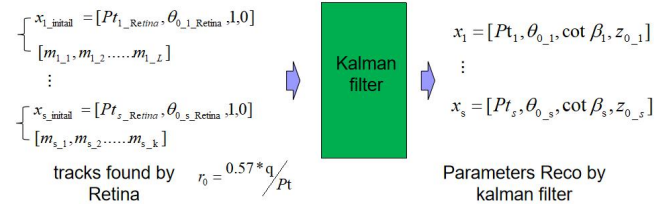

Fig.9. The diagram to show the input and output of kalman filter for track fitting application in tracker detector.

So after Iterative Retina we add the kalman filter to reconstruct the parameters of the tracks found by Iterative Retina with higher precision. Kalman filter is widely use as the parameter reconstruction algorithm for many applications [9]. Figure 9 shows the data flow of our kalman filter implementation after Iterative Retina. We use the hit data from each track that is found by Iterative Retina as the input measurement vector $m_t$ of kalman filter and the two parameters $\theta_0$ and Pt that are reconstructed by Retina as the initial state vector ($x_{1\_initial}$ for the 1st track and $x_{s\_initial}$ for the sth track) of kalman filter, here in the initial state vector $x_{initial}$ we set other two parameters to 1 and 0 for the initial $\cot\beta$ and $z_0$ respectively. To reconstruct four parameters for each track (for example the 1st track in Figure 9), we first input one hit data $m_1$ that belong to the 1st track in kalman filter, after calculation by kalman filter the parameter state x will be update from the current state at current time $x_{t|t}$ to the next state at next time $x_{t+1|t+1}$. Then we input the rest of hit data one by one and do the same procedure over and over again, we can get the final state $x_1$ for the 1st track in the end. This state vector $x_1$ is output as the parameters reconstruct result of the 1st track of the kalman filter. Therefore, before kalman filter we just reconstruct the 2-D parameter ($\theta_0$,Pt) with Retina and after kalman filter one can reconstruct all the 3-D parameter ($\theta_0$,Pt,$\cot\beta$,$z_0$) for each track.

Step 1: $x_{t+1|t} = x_{t|t}$ ①   $P_{t+1|t} = P_{t|t}$ ②

Step 2: $x_{t+1|t+1} = x_{t+1|t} + W_t v_t$ ③   $P_{t+1|t+1} = P_{t+1|t} - W_t H_t P_{t+1|t}$ ④

$x_{t+1|t+1} = x_{t|t} + W_t v_t$ ⑤   $P_{t+1|t+1} = P_{t|t} - W_t S_t W_t^T$ ⑥

Where: $S_t = H_t P_{t|t} H_t^T + R_t$ ⑦   $W_t = P_{t|t} H_t^T S_t^{-1}$ ⑧

$v_t = m_t - H_t x_{t|t}$ ⑨

Fig.10. The calculation steps and related formula to figure out the next state at next time $x_{t+1|t+1}$ using the current state at current time $x_{t|t}$ and the measurement $m_t$.

The Figure 10 represents the formulas and calculation steps to use the current state at current time $x_{t|t}$ and the measurement $m_t$ to figure out the next state at next time $x_{t+1|t+1}$ in kalman filter. At the beginning we have the $x_{t|t}$ and $P_{t|t}$, where $P_{t|t}$ is the covariance matrix for current state at current time. In first step we use formula ① and ② in Fig. 10 to get the state matrix $x_{t|t+1}$ and covariance matrix $P_{t|t+1}$ for current state at next time, and then in second step we use formula ③ and ④ in Fig. 10 to get the state matrix $x_{t+1|t+1}$ and covariance matrix $P_{t+1|t+1}$ for next state at next time. So in the end the relationship between $x_{t|t}$ and $x_{t+1|t+1}$ is represent by formula ⑤ in Fig. 10, and the relationship between $P_{t|t}$ and $P_{t+1|t+1}$ is represent by formula ⑥ in Fig. 10 as well. Moreover, the formula ⑦⑧⑨ in Fig.10 gives the intermediate computer calculation steps for kalman filter, where $H_t$ is the transform matrix and $R_t$ is the measurement covariance matrix.

Transform matrix:
$$H_t = \frac{\partial m}{\partial x} = \begin{bmatrix} r & 1 & 0 & 0 \\ 0 & 0 & r & 1 \end{bmatrix} \quad ①$$

Initial covariance matrix:
$$P_{0|0} = \begin{bmatrix} \sigma_a^2 & \sigma_{ab} & 0 & 0 \\ \sigma_{ab} & \sigma_b^2 & 0 & 0 \\ 0 & 0 & \sigma_c^2 & \sigma_{cd} \\ 0 & 0 & \sigma_{cd} & \sigma_d^2 \end{bmatrix} \quad ②$$

Measurement covariance matrix:
$$R_t = \begin{bmatrix} \sigma_\phi^2 & 0 \\ 0 & \sigma_z^2 \end{bmatrix} \quad \sigma_\phi^2 = \left(\frac{1}{\sqrt{12}}\frac{p}{r}\right)^2 \sigma_z^2 = \left(\frac{1.5625l}{\sqrt{12}}\right)^2 \quad ③$$

Fig.11. Related constant matrices in kalman filter for track reconstructing.

Besides, Figure 11 shows the related constant matrices used in kalman filter calculation. Formula ① in Fig. 11 gives the definition of transform matrix $H_t$, where r is the radius of the hits in x-y plane. Formula ② in Fig. 11 gives the initial covariance matrix in kalman filter for our application, in the formula we can set $\sigma_{ab}$ and $\sigma_{cd}$ equal to 0, the value of $\sigma_a^2$ and $\sigma_b^2$ are $1/\sqrt{12}$ of the corresponding Retina cell parameter width, the value of $\sigma_c^2$ is $1/\sqrt{12}$ of the corresponding sector width we focus on, and the value of $\sigma_d^2$ is equal to the beam spot width 5cm. Formula ③ in Fig. 11 gives the definition of the measurement covariance matrix $R_t$. It indicates the uncertainties on the hit measurements in tracker detector. Where here p is the strip pitch, and l is the strip length: 5 cm in the 2S modules; and 1.5 mm in the PS modules.

## VI. RESULTS ANALYSIS AND CONCLUSION

We test the Iterative Retina in both software and firmware, and with the kalman filter only in software. During the test we have adjusted the parameters of Iterative Retina and kalman filter to appropriate state, and here we represent the final results for our algorithm including the efficiency and purity of Iterative Retina and the parameter reconstruction resolution of both Iterative Retina and kalman filter.

TABLE II
THE EFFICIENCY AND PURITY OF ITERATIVE RETINA IN CASE OF THREE CONFIGURATIONS.

| Case | Ghost track | Mismatched track | Matched track | Duplicated tracks | Total tracks in TTbar event | Efficiency | Purity |
|---|---|---|---|---|---|---|---|
| Loop-48 | 460 | 125 | 1305 | 2362 | 3792 | 96.71 | 88.85 |
| Loop-80 | 167 | 128 | 2312 | 1352 | 3792 | 96.62 | 95.64 |
| Loop-200 | 40 | 238 | 2825 | 729 | 3792 | 94.78 | 98.89 |

The table 2 shows the efficiency and purity of Iterative Retina . As shown in the table the efficiency and purity in these three cases are very closely and both of them are over 90%. So with the Iterative Retina most of the track in t-tbar event can be identified even in case of the reduced scanning. This is due to the suitable thresholds setting in all three cases. With the increased granularity in Retina algorithm the purity increases, because the condition to reach the threshold is more stringent. Therefore, there are less ghost tracks found in that case. On the other hand, the probability of unrecognized tracks will increase (see the number of mismatched tracks in table 2), and the efficiency of the algorithm will be slightly reduced with the increasing of the granularity of Iterative Retina.

Figure 12 and Figure 13 represent the distribution of reconstruct Pt and $\theta_0$ versus input Pt and $\theta_0$, respectively. The left plots show the performance after Iterative Retina and the right plots the same after the kalman filter. From Figure 12 we can see that the precision on the reconstructed Pt decreases at high Pt in all iterative configurations and increases with higher granularities. From Figure 13 we can see that the precision for $\theta_0$ reconstruction by Iterative Retina improves at higher granularity and that the performance are the same for the different direction along $\theta$ in x-y plane. For both track parameters, the kalman filter improves further the resolution whatever the Iterative Retina granularity.

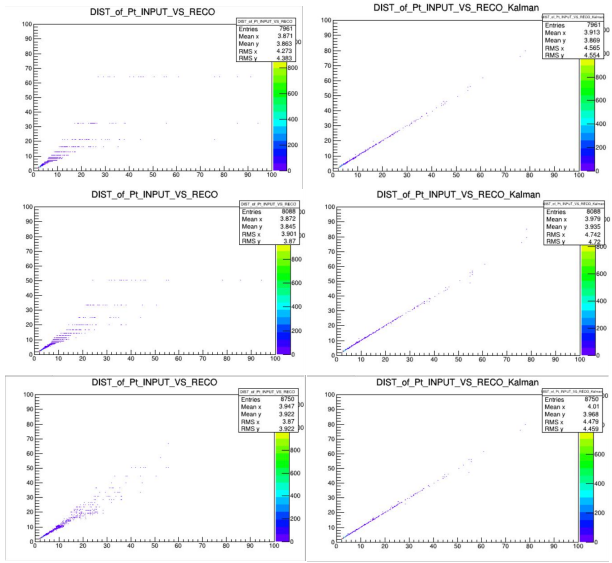

Fig.12. The distribution of reconstruct Pt versus input Pt in three configuration, from up to bottom are Loop-48 Loop-80 and Loop-200 respectively. The 3 plots in left are the results only use Iterative Retina and the 3 plots in right are the results with the kalman filter. The unit of x and y axis are both GeV/c

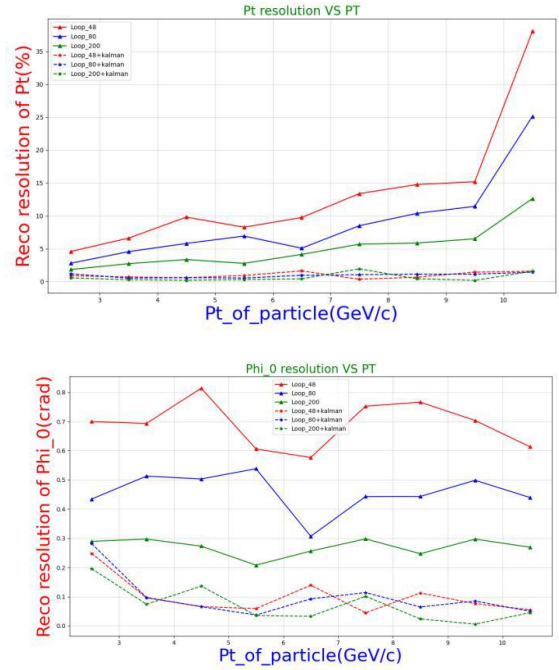

Fig.14. Pt and $\theta_0$ reconstruct resolution in 6 cases. The plot shown in top is the Pt reconstruct resolution versus input Pt and the plot shown in bottom is the reconstruct resolution versus input Pt.

Finally Figure 14 shows the resolution on Pt and $\theta_0$ as a function of the particle Pt before and after the kalman filter. As expected the resolution improves with Retina granularity, and the KF significantly improves it further, even in the case of low Retina granularity, which seems to indicate that we can relax Retina granularity, which means reducing the FPGA resource usage or the processing time of the Retina step. Our next objectives are now a complete and optimized implementation on FPGA.

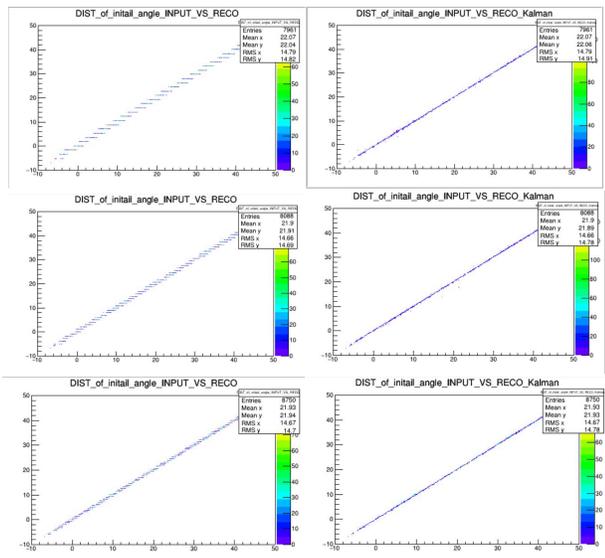

Fig.13. The distribution of reconstruct initial angle $\theta_0$ versus input $\theta_0$ in three configuration, from up to bottom are Loop-48 Loop-80 and Loop-200 respectively. The 3 plots in left are the results only use Iterative Retina and the 3 plots in right are the results with the kalman filter. The unit of x and y axis are both centi-rad(crad)